\documentclass[preprint,aps]{revtex4}

\newcommand {\bc}{\begin{center}}
\newcommand {\ec}{\end{center}}
\newcommand {\bea}{\begin{eqnarray}}
\newcommand {\eea}{\end{eqnarray}}
\newcommand {\be}{\begin{equation}}
\newcommand {\ee}{\end{equation}}

\def\lsim{\mathrel{\rlap{\lower4pt\hbox{\hskip1pt$\sim$}}
    \raise1pt\hbox{$<$}}}               
\def\gsim{\mathrel{\rlap{\lower4pt\hbox{\hskip1pt$\sim$}}
    \raise1pt\hbox{$>$}}}                
\usepackage{graphicx}

\begin{document}


\title{Shear viscosity and damping of collective modes in a two-dimensional 
Fermi gas}

\author{Thomas~Sch\"afer}

\affiliation{Department of Physics, North Carolina State University,
Raleigh, NC 27695}

\begin{abstract}
We compute the shear viscosity of a two dimensional Fermi gas 
interacting via a short range potential with scattering length 
$a_{2d}$ in kinetic theory. We find that classical kinetic theory 
predicts that the shear viscosity to entropy density ratio of a 
strongly interacting two dimensional gas is comparable to that of 
the three dimensional unitary gas. We apply our results to the damping 
of collective modes of a trapped Fermi gas, and compare to experimental 
data recently obtained in E.~Vogt et al., arXiv:1111.1173. 
\end{abstract}

\maketitle

\section{Introduction}
\label{sec_intro}

 The study of transport properties of strongly interacting, scale 
invariant or approximately scale invariant fluids has led to many 
recent discoveries that connect the physics of cold atomic gases, 
properties of the quark gluon plasma, and quantum gravity 
\cite{Schafer:2009dj}. Nearly ideal hydrodynamic flow in cold atomic
gases was observed in the expansion of a dilute Fermi at unitarity 
\cite{OHara:2002}, and similar results were observed in heavy ion 
collisions at the relativistic heavy ion collider (RHIC) \cite{rhic:2005}. 
Recent analyses show that both the quark gluon plasma and the dilute 
Fermi gas at unitarity are characterized by a shear viscosity to 
entropy density ratio $\eta/s\lsim 0.5\, \hbar/k_B$ 
\cite{Dusling:2007gi,Romatschke:2007mq,Schafer:2007pr,Cao:2010}.
This result is close to the value $\eta/s=\hbar/(4\pi k_B)$ which was
found in the strong coupling limit of a large class of field theories 
that can be analyzed using the AdS/CFT correspondence 
\cite{Policastro:2001yc,Kovtun:2004de}. 

 The AdS/CFT result is independent of the dimensionality of the 
fluid, and it is interesting to study whether nearly perfect fluidity 
can be observed in two-dimensional fluids. It was suggested, for 
example, that electrons in graphene might behave as a nearly perfect 
fluid \cite{Mueller:2009}. Recently, a group at the Cavendish Laboratory 
investigated the damping of collective modes in a cold atomic Fermi gas 
tightly confined in one direction \cite{Vogt:2011}. Vogt et al.~determined 
the damping constant as a function of $T/T_F$  in the range $T/T_F=
(0.3-0.8)$, and for different interaction strengths $\log(k_Fa_{2d})=
(2.7-42)$. Here, $T/T_F$ is the temperature in units of the Fermi 
temperature, $k_F$ is the Fermi momentum, and $a_{2d}$ is the two-dimensional 
scattering length. In the present work we compare these results with the 
predictions of kinetic theory. Formally, kinetic theory is reliable in the 
limit of high temperature, $T\gg T_F$, or in the case of weak interactions,
$K_Fa_{2d}\gg 1$. In the case of the three dimensional Fermi gas at unitarity 
it was observed that the range of applicability of kinetic theory is larger 
than one might expect, extending down to $T\sim 0.4\,T_F$ 
\cite{Massignan:2004,Bruun:2007,Schaefer:2009px}.

\section{Kinetic theory}
\label{sec_boltz}

 The viscous stress tensor in hydrodynamics is given by $\delta \Pi_{ij}
=-\eta\sigma_{ij}-\zeta\delta_{ij}\langle\sigma\rangle$ with 
\be 
\label{sig_def}
\sigma_{ij}=\partial_iv_j+\partial_jv_i-\frac{2}{d}\, \delta_{ij}
   \partial_k v_k\, , 
\ee
and $\langle\sigma\rangle = \partial_kv_k$. Here, $v_i$ is the flow 
velocity and $d=2,3,\ldots$ is the number of spatial dimensions. 
We will determine $\eta$ by matching the hydrodynamic result to 
kinetic theory. The stress tensor in kinetic theory is given by
\be
\label{del_pi}
\delta\Pi_{ij} = \nu \int d\Gamma_p \, \frac{p_ip_j}{m} 
 \, \delta f_p\, , 
\ee 
where $\nu$ is the number of degrees of freedom ($\nu=2$ for a 
two-component Fermi gas), $d\Gamma_p=\frac{d^dp}{(2\pi)^d}$ is the 
volume element in momentum space, and $\delta f_p$ is the off-equilibrium 
correction to the distribution function. We will use the ansatz
\be 
\label{del_f}
\delta f_p = f^0_p - \frac{f^0_p}{T} \, \chi_{ij}(p) \sigma_{ij}\, ,
\hspace{1cm}
\chi_{ij}(p)=p_{ij} \chi\left(\frac{p^2}{2mT}\right)\, , 
\ee
where $f^0_p$ is the classical equilibrium distribution function and 
$p_{ij}=p_ip_j-\frac{1}{d}\delta_{ij}p^2$. We will study the role
of quantum statistics below. We compute $\delta f_p$ by solving the 
Boltzmann equation for fermions with dispersion relation $E_p=\frac{p^2}
{2m}$ subject to elastic two-body scattering. At this level of approximation 
the bulk viscosity vanishes. This is the correct result for 3d fermions at 
unitarity \cite{Son:2005tj,Enss:2010qh,Castin:2011}, but the bulk viscosity 
is expected to be non-zero for 3d fermions away from unitarity, and for 
2d fermions at any value of the scattering length. The dependence of 
the integral $\int d\omega\,\zeta(\omega)$ on $(k_Fa)^{-1}$ is constrained 
by sum rules \cite{Taylor:2010ju,Goldberger:2011hh,Hofmann:2011qs}, but 
the bulk viscosity at zero frequency has not been determined. Vogt 
et al.~measured the damping of a 2-d quadrupole mode \cite{Vogt:2011}, 
which is not sensitive to bulk viscosity. 

 Matching the kinetic theory expression for $\delta\Pi_{ij}$ to 
hydrodynamics we get 
\be 
\label{eta_kin}
 \eta = \frac{2\nu}{(d-1)(d+2)}\,
        \frac{1}{mT}\, \langle p_{ij}|\chi_{ij}\rangle\, , 
\ee
where we have defined the inner product $\langle a|b\rangle = \int 
d\Gamma_p\,f^0_p a(p)b(p)$. The function $\chi_{ij}(p)$ is determined 
by the linearized Boltzmann equation 
\be
\label{lin_be}
 \frac{1}{2m} \, |p_{ij}\rangle  = C |\chi_{ij}\rangle \, .
\ee
Here $C$ is the linearized collision operator $C|\chi_{ij}\rangle 
=|C[\chi_{ij}]\rangle$ with 
\bea
 C[\chi_{ij}(p_1)] &=& \prod_{i=2}^4\left(\int d\Gamma_i\right)
 f^0(p_2) 
 (2\pi)^{d+1}\delta^d(P-P')\delta(E-E')\left|{\cal T}\right|^2  \nonumber \\
\label{C_def}
 & & \hspace{2.5cm}\mbox{}\cdot
\left[\chi_{ij}(p_1)+\chi_{ij}(p_2)-\chi_{ij}(p_3)-\chi_{ij}(p_4)\right]
\, , 
\eea
where ${\cal T}$ is the T-matrix for elastic two-body scattering 
$12\rightarrow 34$. We have also defined $p_{1,2}=\frac{P}{2}\pm q$, 
$p_{3,4}=\frac{P'}{2}\pm q'$, $E=E_{p_1}+E_{p_2}$ and $E'=E_{p_3}+E_{p_4}$. 
Given the T-matrix we can determine $\chi_{ij}$ from equ.~(\ref{lin_be}) 
and then compute the shear viscosity using equ.~(\ref{eta_kin}). In 
practice it is useful to reformulate the calculation as a variational 
problem. The shear viscosity can be written as  
\be 
\label{eta_var}
\eta = \frac{\nu}{(d-1)(d+2)}\,\frac{1}{m^2T} \,
  \frac{\langle\chi_{ij}|p_{ij}\rangle^2}
   {\langle \chi_{ij}|C|\chi_{ij}\rangle} \, . 
\ee
The equivalence of this result and the previous expression given 
in equ.~(\ref{eta_kin}) follows from the linearized Boltzmann equation. 
The result is variational in the sense that for a trial function 
$\chi^{\it var}_{ij}$ equ.~(\ref{eta_var}) provides a lower bound on 
the shear viscosity. The exact solution of the linearized Boltzmann 
equation can be found by maximizing equ.~(\ref{eta_var}). In the three
dimensional case it is known that the quadratic ansatz $\chi_{ij}=p_{ij}$ 
is an excellent solution, providing results for the shear viscosity that 
are accurate to $2\%$ \cite{Bruun:2006}. We will see that despite the 
different structure of the scattering amplitudes in two and three
dimensions the matrix elements of the collision operator are very 
similar. We will therefore use the trial function $\chi_{ij}=p_{ij}$.

 In two dimensions the scattering matrix for elastic scattering 
mediated by a short range potential is given by \cite{Randeria:1990}
\be 
{\cal T} = \frac{4\pi}{m}\frac{1}{-\log(q^2a^2_{2d})+i\pi}\, ,
\ee
where $a_{2d}$ is the two-dimensional scattering length. The cross 
section is $\frac{d\sigma}{d\Omega}=\frac{m^2}{4q}|{\cal T}|^2$. 
The matrix element of the linearized collision operator can be 
reduced to a one-dimensional integral. We find
\be
\label{cCc_2d}
 \langle \chi_{ij}|C|\chi_{ij}\rangle = 4 T (mT)^3
 \int_0^\infty dx \, \frac{x^5e^{-x^2}}{\log^2(x^2T/T_{a,2d})+\pi^2}\, ,
\ee
where we have defined $T_{a,2d}=1/(ma_{2d}^2)$.  The integral in 
equ.~(\ref{cCc_2d}) can be computed using the saddle point approximating.
This amounts to replacing the term $x^2$ in the denominator by $5/2$. The 
final result for the shear viscosity is
\be 
\label{eta_2d}
\eta_{2d} = \frac{mT}{2\pi^2}
 \left(\left[\log\left(\frac{5T}{2T_{a,2d}}\right)\right]^2
+\pi^2\right)\, ,
\ee
where we have set $\nu$, the number of spin states, equal to two. We 
can use this results to compute the dimensionless quantities $\eta/n$ 
and $\eta/s$. We find
\be 
\label{eta_n}
\frac{\eta_{2d}}{n} =\frac{\pi}{2}\left(\frac{T}{T_F^{\it loc}}\right)
 \left( 1 +\frac{1}{\pi^2}\left[\log\left(\frac{5T}{2T_{a,2d}}\right)
 \right]^2\right)\, ,
\ee
where $T^{\it loc}_F=(k^{\it loc}_F)^2/(2m)$ is a function of the local 
Fermi momentum, $k_F^{\it loc}=(2\pi n)^{1/2}$. The entropy per particle 
is $s/n=\log(T/T_F^{\it loc})+2$. 

 It is instructive to compare these expressions to the analogous formulas 
in three dimensions. The T-matrix is 
\be
{\cal T}=\frac{4\pi}{m}\frac{1}{-a_{3d}^{-1}+iq}\, , 
\ee
and the cross section is $\frac{d\sigma}{d\Omega}=\frac{m^2}{16\pi^2}|
{\cal T}|^2$. The collision integral is 
\be
\label{cCc_3d}
 \langle \chi_{ij}|C|\chi_{ij}\rangle = 
\frac{16m^{7/2}T^{9/2}}{3\pi^{5/2}}
 \int_0^\infty dx \, \frac{x^5e^{-x^2}}{1+T_{a,3d}/(x^2T)}\, ,
\ee
where $T_{a,3d}=1/(ma_{3d}^2)$. At unitarity, $T_{3d}\to\infty$, the integrand 
differs from the result in two dimensions only by logarithmic terms. The
shear viscosity at unitarity is
\be 
\label{eta_3d}
\eta_{3d} = \frac{15}{32\sqrt{\pi}} \, (mT)^{3/2}\, . 
\ee
In the limit $T_{a,3d}/T\gg 1$ we find $\eta_{3d}=5(mT)^{1/2}/(32\sqrt{\pi}
a^2)$. The three dimensions the density is $n=(k_F^{\it loc})^3/(3\pi^2)$, 
and the shear viscosity to density ratio is 
\be 
\frac{\eta_{3d}}{n} = \frac{45\pi^{3/2}}{64\sqrt{2}}\, 
  \left(\frac{T}{T_F^{\it loc}}\right)^{3/2}\, .
\ee
Finally, the entropy per particle is $s/n=\frac{3}{2}
\log(\pi T/T_F^{\it loc})+ \log(3/4)+5/2$. 

\begin{figure}[t]
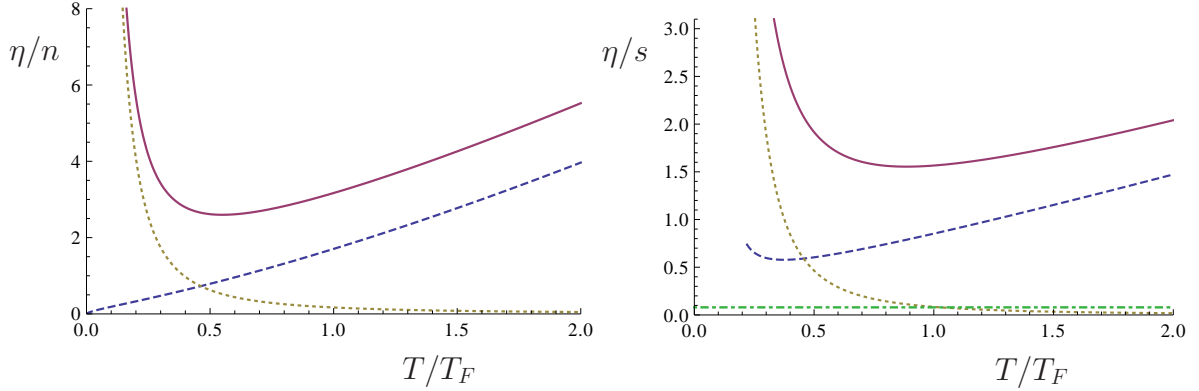

\bc\includegraphics[width=7.75cm]{eta_n_cl_qu_txt.epsi}
\includegraphics[width=7.75cm]{eta_s_cl_qu_txt.epsi}
\ec
\caption{\label{fig_eta_n}
The left panel shows the viscosity to density ratio $\eta/n$ as a function 
of $T/T_F$ for a two-dimensional Fermi gas with $(k_Fa)^2=2$. Here, 
$T_F=k_F^2/(2m)$ and $k_F=(2\pi n)^{1/2}$ characterize the homogeneous 
Fermi gas. The solid line includes the effects of quantum statistics, 
the dashed line shows the high temperature limit given in equ.~(\ref{eta_n}),
and the dotted line shows the low temperature limit. The right panel 
displays the shear viscosity to entropy density ratio. The dash-dotted 
line shows the proposed bound $\eta/s=1/(4\pi)$.}   
\end{figure}

 The results in two dimensions are plotted as the blue dashed lines in 
Fig.~\ref{fig_eta_n}. We have chosen $(k_Fa_{2d})^2=2$, which means 
that the two body binding energy $E_B=1/(ma_{2d}^2)$ is equal to the 
Fermi energy. This corresponds to the BEC/BCS crossover regime. We 
observe that for $T/T_F^{\it loc}\lsim 0.5$ the shear viscosity to 
entropy density ratio reaches $\eta_{2d}/s\simeq 0.5$, comparable to the 
result for the three dimensional Fermi gas at unitarity. In this regime 
kinetic theory is not reliable -- effects due to quantum statistics, 
correlations and fluctuations are likely to play a role. Quantum 
statistics can be included straightforwardly in the kinetic theory
calculation by including appropriate statistical factors in 
equ.~(\ref{del_pi},\ref{del_f}) and (\ref{C_def}). The result is 
shown as the solid line in Fig.~\ref{fig_eta_n}. Pauli blocking 
suppresses the scattering matrix element and leads to $\eta_{2d}/n
\sim (T_F^{\it loc}/T)^2[\log(T_F^{\it loc}/T)]^2$ as $T\to 0$. This 
result is expected from Landau Fermi liquid theory \cite{Novikov:2006}.
We observe that in two dimension the effect of Pauli blocking is quite
large, but we also emphasize that at strong coupling the inclusion of 
quantum statistics is not necessarily an improvement over the classical 
calculation. In the case of thermodynamic quantities, like the second
Virial coefficient, it is well known that effects of quantum statistics 
appear at the same order in $T/T_F$ as higher order interaction terms. 
A similar effect is seen in the many body $T$-matrix calculation of 
the shear viscosity of the three dimensional gas at unitarity by Enss 
et al.~\cite{Enss:2010qh}. These authors include pairing correlations
and vertex corrections in addition to the effects of quantum statistics.
They find that the shear viscosity to entropy density ratio remains 
very close to the classical result even in the very degenerate regime  
$T\sim (0.2-0.5) \,T_F$.

\section{Damping of collective modes in a trapped gas}
\label{sec_relax}

 In hydrodynamics the damping of collective modes is governed by 
the rate of energy dissipation
\be 
\label{e_diss}
\dot{E} = -\frac{1}{2} \int d^3x\, \eta(x) \left( \sigma_{ij}\right)^2\, , 
\ee
where we have neglected bulk viscosity and assumed that the system 
remains isothermal (so that heat conductivity can be neglected). For 
simple modes like the quadrupole oscillation studied by Vogt et 
al.~the velocity field is linear in the coordinates and the stress
tensor is spatially constant. In this case the decay rate is sensitive
to the spatial integral of $\eta(x)$. On dimensional ground we can 
write the viscosity of the homogeneous system as $\eta=n\alpha_n(T/
T^{\it loc}_F,k^{\it loc}_Fa)$. The spatial integral over $\eta(x)$ 
can then be written as $N\langle\alpha_n\rangle$, where $N$ is the 
total number of particles and $\langle\alpha_n\rangle$ is the value 
of $\alpha_n$ averaged over the density distribution of the cloud. 
In the hydrodynamic regime measurements of the damping constant
of collective modes can therefore be interpreted as measurements of 
$\langle\alpha_n\rangle$. 

 The difficulty with this approach is that in kinetic theory $\eta=
n\alpha_n$ is independent of the density and the spatial average 
$\langle\alpha_n\rangle$ is formally infinite. Physically, this 
problem is related to the fact that for any finite collective mode
frequency hydrodynamics cannot be applicable in the dilute corona 
of the cloud, so that the integral in equ.~(\ref{e_diss}) has to be 
cut off at low density \cite{Schaefer:2009px}. In kinetic theory 
this can be done by taking into account the frequency dependence
of the shear viscosity
\be 
\label{eta_om}
\eta(\omega) = \frac{\eta(0)}{1+\tau_R^2\omega^2}\, , 
\ee
where $\tau_R$ is the viscous relaxation time, which is the time 
it takes for the stress tensor to relax to the Navier-Stokes form
$\delta\Pi_{ij}=-\eta(0)\sigma_{ij}$. We will see that the relaxation 
time is inversely proportional to the density, and that the spatial 
integral over $\eta(\omega)$ is therefore finite 
\cite{Bruun:2007,Schaefer:2009px}. 

 The relaxation time can be determined in various ways, for example
by solving the linearized Boltzmann equation in a time-dependent 
velocity field \cite{Nikuni:2004,Bruun:2005}, by computing the 
viscosity spectral function \cite{Chao:2010tk}, or by evaluating the 
relaxation time in second order hydrodynamics \cite{Chao:2011cy}.
The relaxation time is also constrained by viscosity sum
rules \cite{Taylor:2010ju,Goldberger:2011hh,Hofmann:2011qs}.
Using the methods described in \cite{Chao:2010tk} we can show
that in kinetic theory $\eta(\omega)$ satisfies the sum rule
\be
\label{eta_sr}
\frac{1}{\pi} \int d\omega \, \eta(\omega) = \frac{P}{2}\, ,
\ee
where $P$ is the pressure. This sum rule is valid in both two and three 
dimensions. Combining equ.~(\ref{eta_om}) with the viscosity sum rule 
equ.~(\ref{eta_sr}) we get $\tau_R =\eta/P\simeq \eta/(nT)$. 

 We note that the sum rule in equ.~(\ref{eta_sr}) follows from the 
definition of the stress tensor in kinetic theory, see equ.~(\ref{del_pi}). 
If the stress tensor is defined as an operator in the quantum theory
one finds that the spectral function in two dimensions has a $1/\omega$ 
tail at high frequency \cite{Hofmann:2011qs}. The corresponding behavior
in three dimension is $\rho(\omega)\sim 1/\sqrt{\omega}$. This tail does 
not appear in kinetic theory because kinetic theory is an effective theory 
for energies $\omega\lsim T$. In the quantum mechanical sum rule the 
high frequency has to be subtracted. In the high temperature regime, 
$T\gsim T_F$, the conclusion is the same as before: the high frequency 
tail does not contribute to the sum rule, and the width of the transport 
peak is controlled by the relaxation time $\tau_R=\eta/(nT)$
\cite{Enss:2010qh}.

\begin{figure}[t]
\bc\includegraphics[width=10cm]{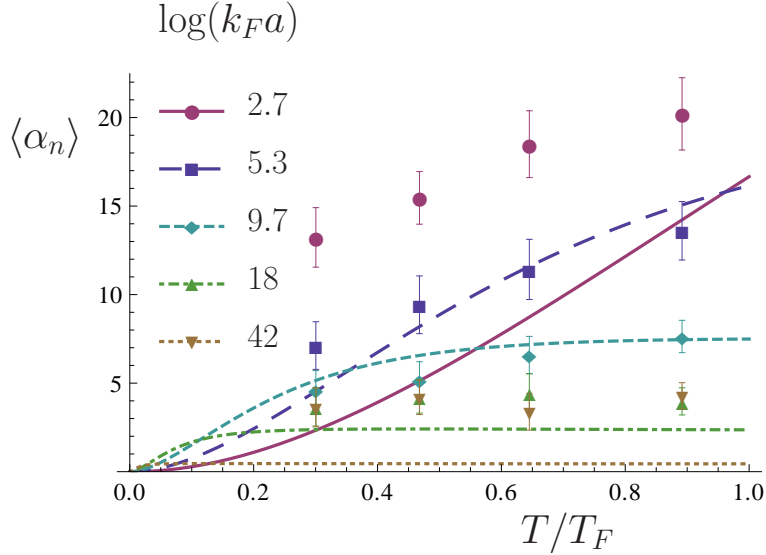}\ec
\caption{\label{fig_alpha_n}
This figure shows the trap average of the shear viscosity to 
density ratio $\alpha_n$ as a function of $T/T_F$ for different 
values of $\log(k_Fa_{2d})$. Here, $T_F\equiv\omega_\perp N^{1/2}$
is the Fermi temperature and $k_F=(2mT_F)^{1/2}$ is the Fermi momentum 
in the trap. The scale is set by $\omega_\perp$, the transverse (two 
dimensional) confinement frequency. We have used $N=4\cdot 10^3$. We compare
our results to the data from Vogt et al.~\cite{Vogt:2011}.}   
\end{figure}

 We can now compute the trap average of $\eta(\omega)$. We will
use the high temperature approximation for the cloud density. 
This is consistent with the classical kinetic calculation of $\eta$, 
and is expected to be a good approximation in the regime $T/T_F\geq
0.3$ studied by Vogt et al. In this limit the density profile
of a 2-dimensional cloud is 
\be
 n(x) = \frac{mT}{2\pi}\left(\frac{T_F}{T}\right)^2
   \exp\left(-\frac{m\omega_\perp^2x^2}{2T}\right)\,
\ee
where $T_F=\omega_\perp N^{1/2}$ is the Fermi temperature of the trapped 
gas. For the 2-dimensional quadrupole mode the frequency is given by 
$\omega=\sqrt{2}\omega_\perp$ \cite{Bulgac:2005,Wen:2007,Klimin:2011}. 
We note that the quadrupole mode is volume conserving, and the frequency 
is independent of the equation of state. We get
\be
\langle\alpha_n\rangle = \frac{1}{2\pi}
  R \left(\frac{T}{T_F}\right)^2
  \log\left[1+\frac{N\pi^2}{2R^2}\left(\frac{T_F}{T}\right)^2\right]\, ,
\hspace{0.5cm}
R = \left[\log\left(\frac{5T}{2T_{a,2d}}\right)\right]^2+\pi^2\, . 
\ee
This result is plotted in Fig.~\ref{fig_alpha_n}. We observe that for 
small values of $\log(k_Fa_{2d})$ and $T/T_F$ the trap average $\langle
\alpha_n\rangle$ grows approximately as $T^2$. This power law can be 
understood as one factor of $T$ arising from the temperature scaling of 
$\eta$, and one factor of $T$ from the inverse density at the center of 
the trap. For larger values of $\log(k_Fa_{2d})$ the growth of the 
relaxation time compensates the growth in $\eta$ and the trap average 
$\langle\alpha_n\rangle$ is only weakly temperature dependent. 

 In Fig.~\ref{fig_alpha_n} we also compare our results to the data 
obtained by Vogt et al.~\cite{Vogt:2011}. We observe that the predicted 
dependence of $\langle\alpha_n\rangle$ on $T/T_F$ and $\log(k_Fa_{2d})$ is 
in qualitatively agreement with the data. The theoretical predictions
are in quantitative agreement with the data for $\log(k_Fa_{2d})= 5.3$ 
and $9.7$. The disagreement between theory and data for $\log(k_Fa_{2d})
= 2.7$ is somewhat puzzling, because this value of $\log(k_Fa_{2d})$ 
corresponds to a more strongly interacting fluid, and we would expect
hydrodynamics to work better. Of course, the kinetic theory calculation 
of the shear viscosity might break down at strong coupling and $T/T_F
\lsim 1$. Another possible issue is that the experimental analysis 
used a free Fermi gas model to estimate the energy of the mode. At 
strong coupling this approach will tend to overestimate the energy,  
and the extracted trap average $\langle\alpha_n\rangle$ is too large. 
The theory also under-predicts the data for large values of $\log(k_F
a_{2d})$. This is less surprising, because hydrodynamics is expected 
to break down in this regime. 

\section{Outlook}
\label{sec_out}

 The observed qualitative agreement between experiment and the 
predictions of kinetic theory suggests that the shear viscosity 
of the two dimensional Fermi gas can be extracted from measurements 
of the damping of collective modes. In order to do this quantitatively 
a number of effects will have to be studied more carefully. We observe, 
in particular, that for $\log(k_Fa_{2d})\gsim 5$ the measured collective 
mode frequencies are not close to the hydrodynamic predictions. This
implies that dissipative effects are not accurately described by 
the hydrodynamic expression given in equ.~(\ref{e_diss}). A more 
appropriate approach is to treat the collective mode itself in kinetic 
theory. This calculation will also provide an indication whether the
observed damping at large $\log(k_Fa_{2d})$ is related to collisions, 
or other effects that are not taken into account in a kinetic or
hydrodynamic treatment. 

 We note that even though the observed trap averaged values 
of $\langle \alpha_n\rangle$ are on the order of 1 or larger 
the corresponding value of $\eta/s$ at the center of the trap 
could be quite small, on the order of $\eta/s\sim 0.5$, see 
Fig.~\ref{fig_eta_n}. In the interesting regime $T\lsim 0.5\,T_F$ 
classical kinetic theory is not reliable. In two dimensions, in 
particular, correlations and fluctuations are likely to play an 
important role. An important example of a correlation effect
is the pseudo-gap phenomenon which was argued to play an 
important role in the transport behavior of the three dimensional
gas \cite{Guo:2010}. A pseudo-gap has been observed in the 
two dimensional gas in the regime $\log(k_F a_{2d})\lsim 
1$ \cite{Feld:2011}. The phase transition in two dimensions is of 
Berezinsky-Kosterlitz-Thouless (BKT) type, and the two dimensional
Fermi gas may provide a very clean system to study transport properties 
near the BKT transition. It is also known that in two dimensions 
hydrodynamic fluctuations lead to a slow, logarithmic, divergence 
of the shear viscosity with the system size \cite{Kadanoff:2009}.

 Acknowledgments: This work was supported in parts by the US 
Department of Energy grant DE-FG02-03ER41260. We thank Michael
K\"ohl and John Thomas for useful discussions. 


\end{document}